\documentclass[10pt]{iopart}
\usepackage{iopams,natbib}
\bibpunct{(}{)}{;}{a}{}{,}
 
\pdfoutput=0
\usepackage{graphicx}
\newcommand{\rd}{{\rm d}}
\newcommand{\bm}[1]{\mbox{\protect\boldmath$#1$}}
\begin{document}
\title[Magnetic Layers and Neutral Points Near Rotating Black Hole]{Magnetic Layers and Neutral Points Near Rotating Black Hole}
\author{V Karas and O Kop\'a\v{c}ek}
\address{Astronomical Institute, Academy of Sciences, Bo\v{c}n\'{i}~II, CZ-141\,31~Prague, Czech~Republic}
\ead{vladimir.karas@cuni.cz, kopacek@ig.cas.cz}

\begin{abstract}
Magnetic layers are narrow regions where the field direction changes
sharply. They often occur in the association with neutral points of the
magnetic field. We show that an organised field can produce these
structures near a rotating black hole, and we identify them as potential
sites of magnetic reconnection. To that end we study the field lines
affected by the frame-dragging effect, twisting the magnetic structure 
and changing the position of neutral points. We consider oblique fields
in vacuum. We also include the possibility of translational motion of
the black hole which may be relevant when the black hole is ejected from
the system. The model settings apply to the innermost regions around
black holes with the ergosphere dominated by a super-equipartition
magnetic field and loaded with a negligible gas content.
\end{abstract}
\pacs{04.20.-q, 97.60.Lf, 98.62.Js}
\maketitle

\section{Introduction}
Magnetic reconnection takes place while topologically distinct regions 
approach each other and the magnetic field lines change their
connectivity \citep{pf00,som06}. The standard setup involves the
violation of the ideal MHD approximation on the boundary between
neighbouring magnetic domains where the field direction changes rapidly.
The essential question is thus about the formation of antiparallel field
lines and neutral points of the magnetic field. Our objective is to
explore if the reconnection process can be influenced by the strong
gravitational field near a black hole (BH), and whether the black-hole
proximity creates conditions favourable to incite reconnection. To this
end we examine the structure of the magnetic field lines twisted by a 
highly curved spacetime of a rotating BH. 

Astrophysical black holes do not support their own intrinsic magnetic
field. Instead, the field is generated by currents in the surrounding
plasma and brought down to the horizon by accretion. Most likely, the
magnetic field is frozen in an equatorial accretion disc which acts as a
boundary condition. The inward-directed bulk transport of plasma is
often in some kind of interaction with outflows and jets
\citep[e.g.,][]{bbr84,blo93}, and so the models of electromagnetic
acceleration and collimation have been continuously advanced during the
last decade \citep{fg01,khh05,kbv07}. The presence of the central BH is
essential. Observations indicate that the initial acceleration happens
very near the horizon \citep{jbl99,a08}, however, the exact mechanism
and the place of the particle acceleration are still a subject of
investigations.

Simulations as well as theoretical considerations point to the decisive
importance of the magneto-rotational instability operating in the inner
regions of the accreting BH systems \citep{bh98,pb03}. On the other
hand, the origin of highly collimated outflows is easier to understand 
with the help of an organised magnetic field. Indeed, it has been argued
that an organised magnetic field is needed to ensure the outflow
collimation that is seen on larger scales, i.e.\ $\gtrsim10r_{\rm g}$ 
in some accreting BH systems \citep{sn92,mn07}. At the same time the
organised field has consequences for the turbulent dynamo mechanism
\citep{hb04}. 

The role of an ordered component of the magnetic field and whether it
may pervade down to the plunging region is a matter of debate
\citep{k04,mn07}. A distinctive feature here is the general-relativistic
(GR) frame-dragging by the BH rotation which, however, can be concealed
by MHD effects in astrophysically realistic situations \citep{km07}.
Turbulent magnetic fields are thought to be limited to sub-equipartition
values, so the rotating disc can indeed enter into the ergosphere.
Plasma acceleration is then driven not only by the frame-dragging effect
of the black hole but mainly by the differential rotation. Here we
address a related issue, of the BH rotation, from another perspective:
we assume a magnetically dominated system. Distinguishing the impact of
the two mechanisms is quite a delicate task when both are in operation.

We assume that the magnetic pressure dominates the studied region of the
ergosphere and has its origin in currents flowing farther out
\citep{kks06,rl08}. Even though the assumption of an ordered field
component constrains the applicability of our calculation, the model
setup can still be reconciled with the recent simulations in which the
magnetic field pressure outside the main flow exceeds the gas pressure
\citep{bl07,bhk08}. Moreover, in absence of the standard accretion disc,
the system does not have to be aligned with the BH equatorial plane and
the gas chunks can arrive episodically \citep{cm97,cnm08}. In these
circumstances one does not expect the BH and the ordered magnetic field
to have a common symmetry axis (the Bardeen-Petterson effect does not
operate due to the lack of a steady accretion flow). Neither we expect
that the black hole is resting in the centre. The possibility of such
mis-aligned accretion has been observed also in the simulations of
geometrically thick accretion \citep[][the  latter computations reveal a
significant part of the equatorial plane near BH to be devoid of gas
although their consistency is restricted by use of the pseudo-Newtonian 
model]{rfm05}. The possibility of rapid linear motion will be addressed
later in the paper.

We demonstrate that antiparallel field lines can be brought in mutual
contact by the frame-dragging alone. Separatrices and the associated
null points occur roughly at the innermost stable circular orbit (ISCO),
where they can act as a place of particle acceleration. It is
interesting to realise that even the (asymptotically) ordered magnetic
field becomes so entangled by the BH gravity that the conditions for
reconnection exist near the horizon. It is also pertinent to remark that
the present-day techniques almost reach the necessary resolution of the
order of one gravitational radius, at least in the case of the largest
on the sky black hole, i.e.\ Sagittarius A* \citep{dwr08}.

\section{Stretching the magnetic lines by gravitational frame-dragging}
We setup an idealised framework which maximises the action of
frame-dragging due to the BH rotation while suppressing the MHD effects
caused locally due to plasma motions. The gravitational field is
described by the Kerr metric,
\begin{equation}
{\rm d}s^{2} = 
  -\frac{\Delta\Sigma}{A}\,{\rd}t^{2}
   +\frac{\Sigma}{\Delta}\,{\rd}r^{2}
   +\Sigma\,{\rd}\theta^{2}
+\frac{A\sin^2\theta}{\Sigma}\;
   \left({\rd}\phi-\omega\,{\rd}t\right)^{2},
\label{metric}
\end{equation}
where $a$ denotes the specific angular momentum, $|a|\leq1$ ($a=0$ is
for a non-rotating BH, while $a=\pm1$ is for the maximally
co/counter-rotating one), $\Delta=r^{2}-2r+a^{2}$,
$\Sigma=r^{2}+a^{2}\mu^2$, $A=(r^{2}+a^{2})^{2}-{\Delta}a^{2}\sigma^2$,
and $\omega=2ar/A$.\footnote{Hereafter, we use  geometrical units,
$c=G=1$, and scale all quantities by the BH mass $M$ \citep{mtw73}. The
gravitational radius is 
$r_{\rm{}g}=c^{-2}GM\approx4.8\times10^{-7}M_7\,$pc, and the
corresponding light-crossing time-scale
$t_{\rm{}g}=c^{-3}GM\approx49\,M_7\,$sec, where $M_7\equiv
M/(10^7M_\odot)$. We use spheroidal coordinates with
$\mu\equiv\cos\theta$, $\sigma\equiv\sin\theta$.} The outer horizon,
$r\equiv r_+(a)$, is where $\Delta(r)=0$, whereas the ISCO ranges
between $1\leq r_{\rm{}isco}(a)\leq9$ \citep{bpt72}. The presence of
terms $\propto\omega$ in the metric (\ref{metric}) indicates that the GR
frame-dragging operates and affects the motion of particles as well as
the structure of fields \citep[e.g.,][]{kv94,mvs99,bhr07}. This
influence on the accreted matter resembles a rotating viscous medium
forcing the field lines to share the rotational motion, while at the
same time the electric component is generated \citep{p08}.

We assume that the electromagnetic field does not contribute to the 
system gravity, which is correct for every astrophysically realistic
situation. Within a limited volume around the BH, typically of the size
$\ell\simeq10$, the magnetic field lines exhibit a structure
resembling the asymptotically uniform field. The electromagnetic field
is a potential one and can be written as a superposition of two parts:
the aligned component \citep{wal74,kkl75}, plus the asymptotically
perpendicular field \citep{bj80,ag89}. The four-potential is:
\begin{eqnarray}
A_t &\!\!\!=\!\!\!&  
 B_{\parallel}a\left[r\Sigma^{-1}\left(1+\mu^2\right)
 -1\right]
 + B_{\perp}a\Sigma^{-1}\Psi\sigma\mu, \label{mf1} \\
A_r &\!\!\!=\!\!\!& -B_{\perp}(r-1)\sigma\mu\sin\psi, \\
A_{\theta} &\!\!\!=\!\!\!& 
 -B_{\perp}\big[\big(r\sigma^2+\mu^2\big) a\cos\psi 
 +
 \big(r^2\mu^2+\big(a^2-r\big)(\mu^2-\sigma^2)\big)
 \sin\psi\big], \\
A_{\phi} &\!\!\!=\!\!\!& B_{\parallel}\big[
{\textstyle\frac{1}{2}}\big(r^2+a^2\big)
 -a^2r\Sigma^{-1}\big(1+\mu^2\big)\big] \sigma^2 \nonumber \\
 && -B_{\perp}\big[\Delta\cos\psi+\big(r^2+a^2\big)
 \Sigma^{-1}\Psi\big] \sigma\mu, \label{mf4}
\end{eqnarray}
where
$\psi\equiv\phi+a\delta^{-1}\ln\left[\left(r-r_+\right)/\left(r-r_-\right)\right]$
is the Kerr ingoing angular coordinate, $\Psi=r\cos\psi-a\sin\psi$,
$\delta=r_+-r_-$, and $r_{\pm}=1\pm\sqrt{1-a^2}$.

Eqs.~(\ref{mf1})--(\ref{mf4}) describe the electromagnetic field in
vacuum. They were originally derived in the  above-mentioned papers and
further explored in various follow-ups \citep[e.g.,][]{k89}. Even if the
vacuum fields are obviously a simplification as far as the realistic
systems are concerned, they may actually capture some important aspects
of the field outside dense plasma distributions, i.e., away from the
main body of the accretion flow. These properties are likely to persist
as long as the magnetic field is dragged in a (slightly) different pace
than the plasma itself (the case of resistive magneto-hydrodynamics),
although a detailed form of the realistic solution will be different.
One can argue, and notice in the simulations, that whereas the currents
in plasma do create a turbulent field structure in their immediate
neighbourhood, as soon as one looks outside the plasma flow the field
structure becomes simpler and organised on larger scales. 

Most important for our arguments is the fact that the solution
(\ref{mf1})--(\ref{mf4}) includes the interaction of the BH strong
gravity with the electromagnetic field, including the influence of the
frame dragging. A question arises whether the ordered field
configuration can efficiently accelerate the particles from the vicinity
of the horizon. Magnetic reconnection assisted by the gravitational
frame dragging could help to solve this puzzle.

The magnetic field is directed in a general angle with respect to the BH
rotation axis. Only few aspects of these misaligned magnetic fields
have been explored so far \citep[see][]{t00,na07}. The transversal
component is wound up around the horizon and it is {\it not\/} expelled
out of the horizon (i.e., the the transversal magnetic flux across the
horizon does not vanish even in the extremely rotating case;
\citealt{bkl07}). On the other hand, the behaviour of the aligned
component is simpler, the field being gradually expelled out of the
horizon as the BH rotation increases \citep{kkl75}. Conductivity of the
medium changes this property \citep{mg04,km07}, and so non-vacuum fields
are more complicated.

To obtain the physical components of the electromagnetic tensor,
$\bm{F}=2\,\bm{\rd A}$, we project $\bm{F}$ onto the local tetrad,
$\bm{e}_{(a)}$. The appropriate choice of the projection tetrad is the
one attached to a frame in Keplerian orbital motion, which exists for
$r\geq r_{\rm{}isco}(a)$. Below $r_{\rm{}isco}(a)$ the circular motion
is unstable and the matter has to spiral downwards while maintaining
constant angular momentum of $l\equiv l(r_{\rm{}isco})$. Only in the
extremely rotating case is the circular motion possible all the way down
to $r_{\rm{}isco}=1$. The electric and magnetic intensities,
measured by a physical observer, are:
$E_{(a)}=\bm{e}_{(a)}^{\mu}F_{\mu\nu}u^{\nu}$,
$B_{(a)}=\bm{e}_{(a)}^{\mu}F^{*}_{\mu\nu}u^{\nu}$, where
$u^\nu\equiv\bm{e}_{(t)}^{\nu}$ is the observer's four-velocity (the
remaining three basis vectors can be chosen as space-like, mutually
perpendicular vectors). The dependence of the electromagnetic components
on the $\psi$-angle indicates the ever increasing effect of the frame
dragging near the horizon.

Assuming an ordered magnetic field is obviously a crude starting point,
but a sensible one. It approximates the field generated by sources
distant from the BH. The magnetic intensity of the organised component
in real systems is quite uncertain \citep[e.g.,][]{m05}. By analogy
\citep[see][]{mfy01,m06,ebz08} we can expect $\simeq10$ gauss acting on
length-scales of $\simeq10$--$20$ gravitational radii near 
the Galactic centre supermassive BH. 
This allows us to estimate the maximum energy to which a particle can be
accelerated by an equipartition electric field acting along the distance
$\ell$, $E_{\rm{}max}\simeq
10^{18}q_{\rm{}e}(B/10\mbox{G})\,(\ell/r_{\rm{}g})$~eV, where
$q_{\rm{}e}$ is in units of the elementary charge. Naturally, this
qualitative estimate is exceeded if a non-stationary field governs the
acceleration process.

\begin{figure*}[tbh!]
\includegraphics[width=.49\textwidth]{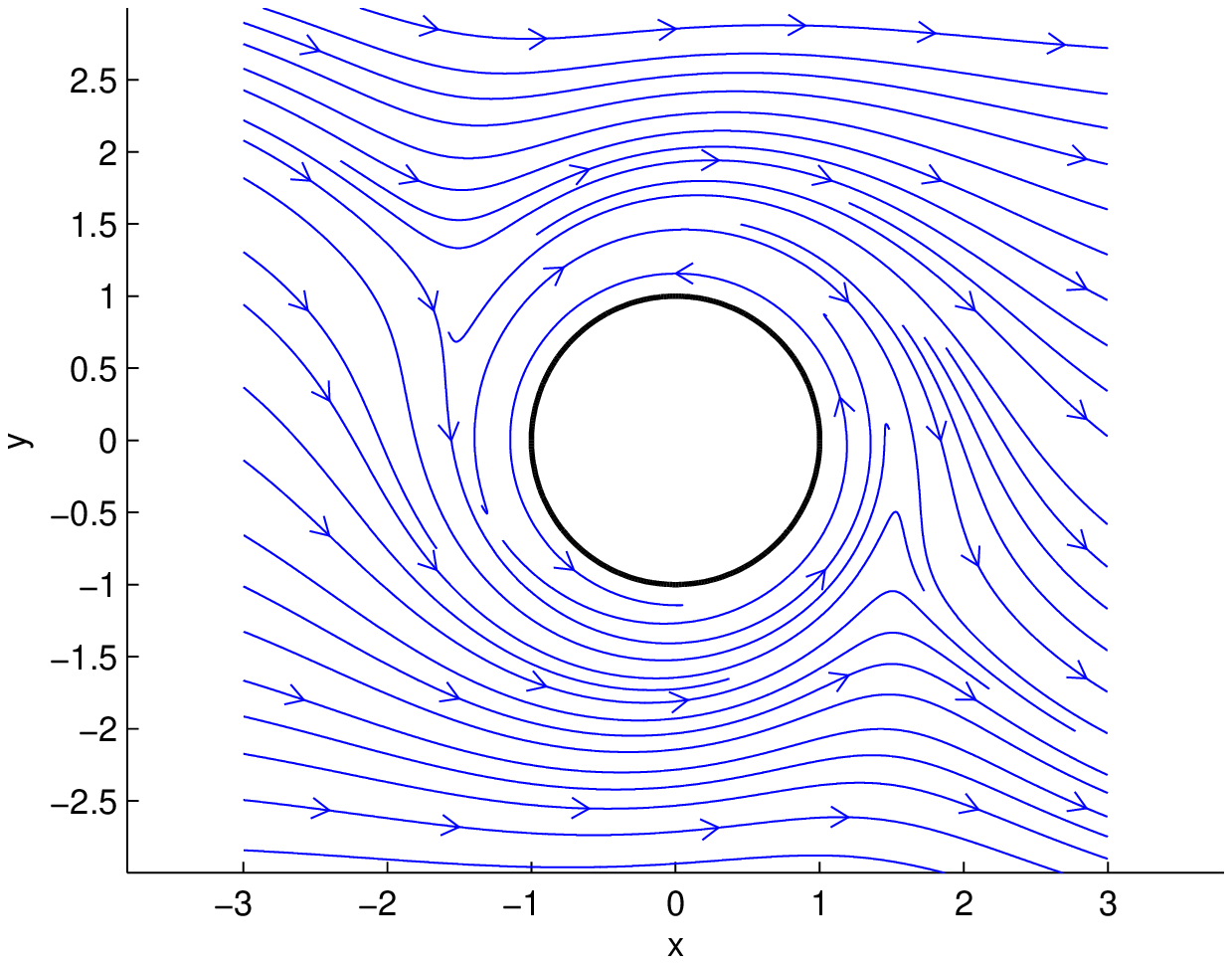}
\includegraphics[width=.49\textwidth]{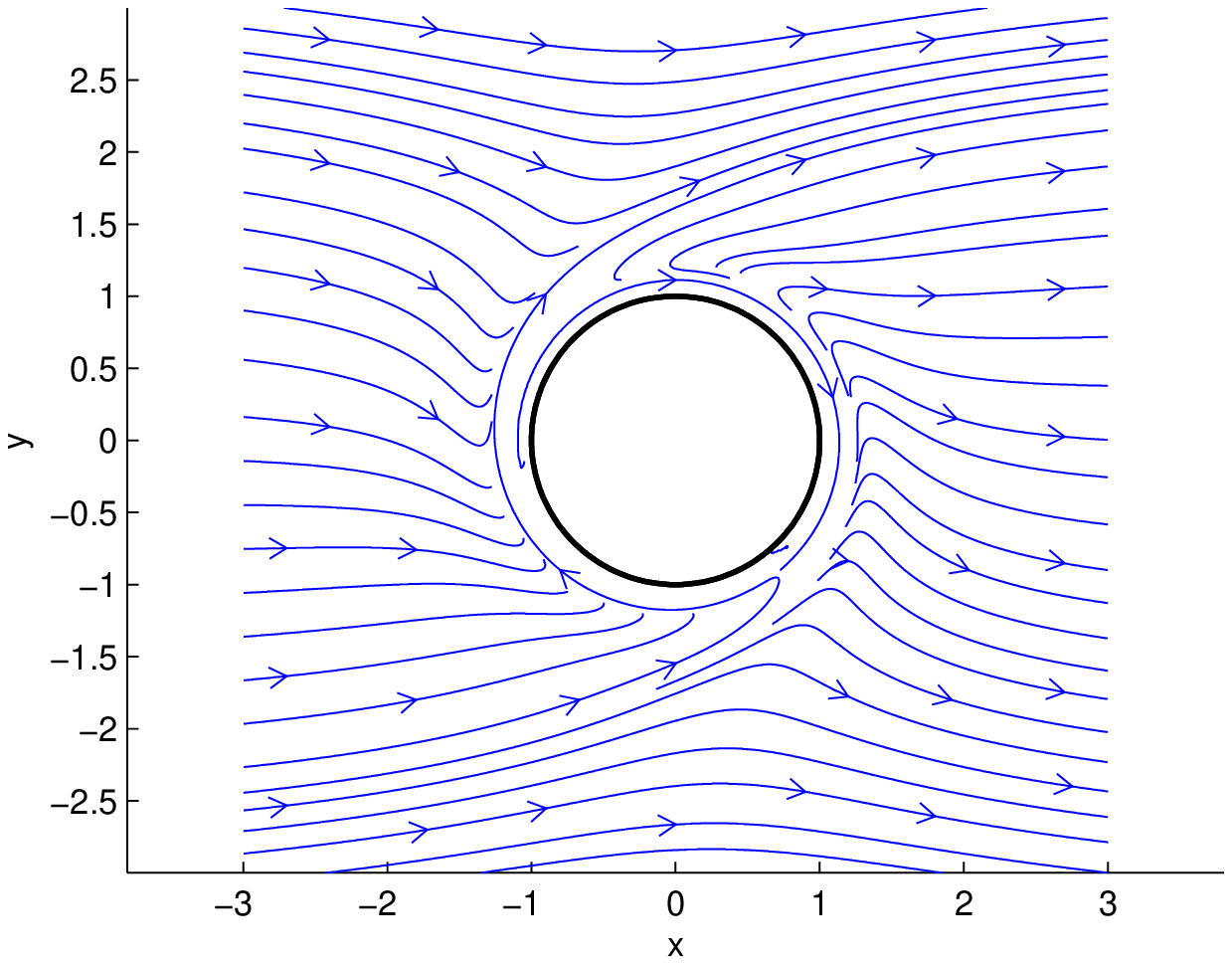}
\caption{The magnetic field lines in the equatorial plane $xy$, 
perpendicular to the rotation axis of an extremely rotating BH 
(coordinates are scaled in units of $r_{\rm{}g}$). The black hole is
resting at the origin; its horizon is denoted by the circle. An
asymptotically uniform magnetic field is directed along the $x$-axis at
large radii and plotted with respect to the physical frame in the
Keplerian orbital motion. Left: the case of a co-rotating frame with
respect to the BH ($a=1$). Right: the counter-rotating case ($a=-1$). 
\label{fig1}}
\end{figure*}

\begin{figure*}[tbh!]
\includegraphics[width=.49\textwidth]{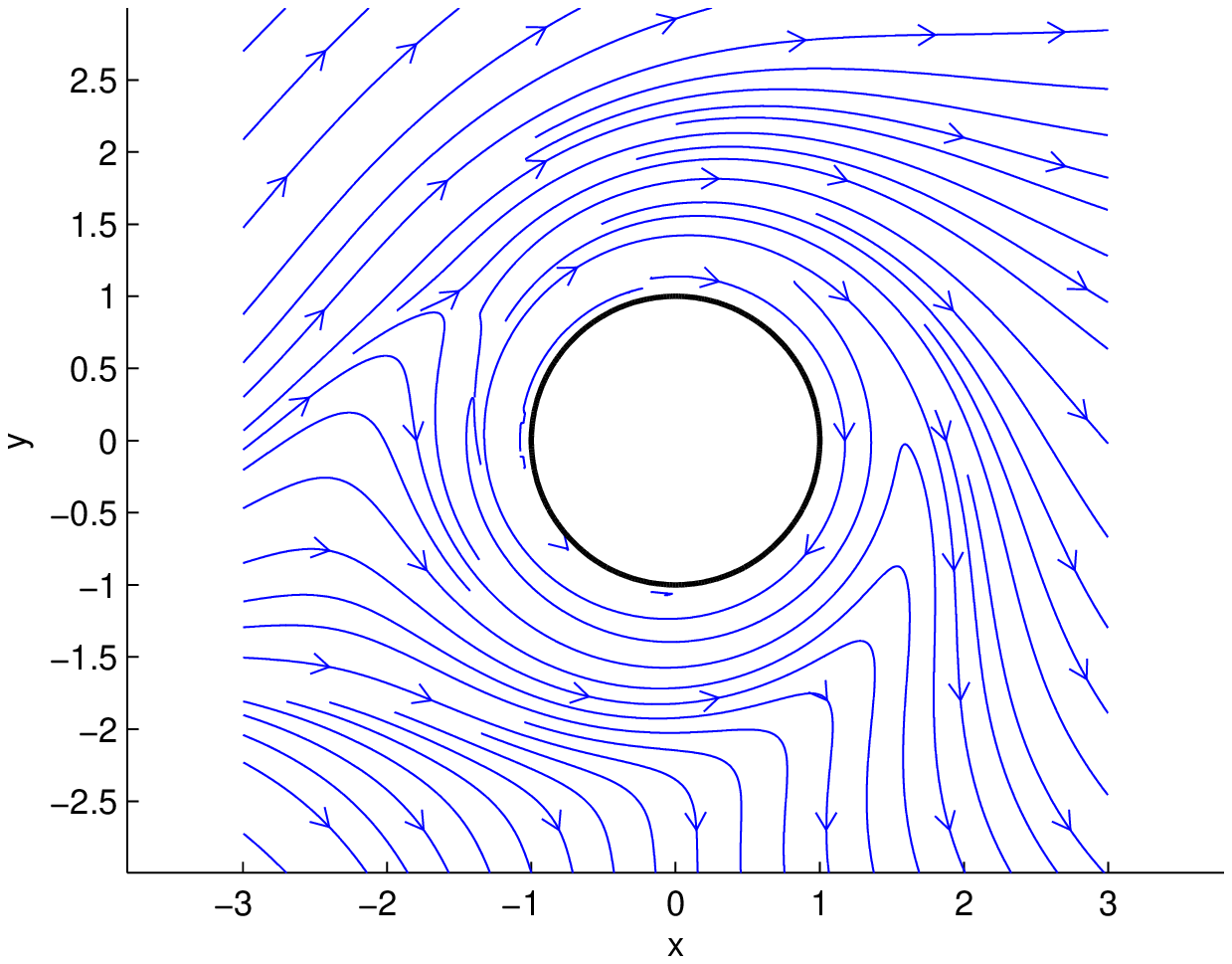}
\includegraphics[width=.49\textwidth]{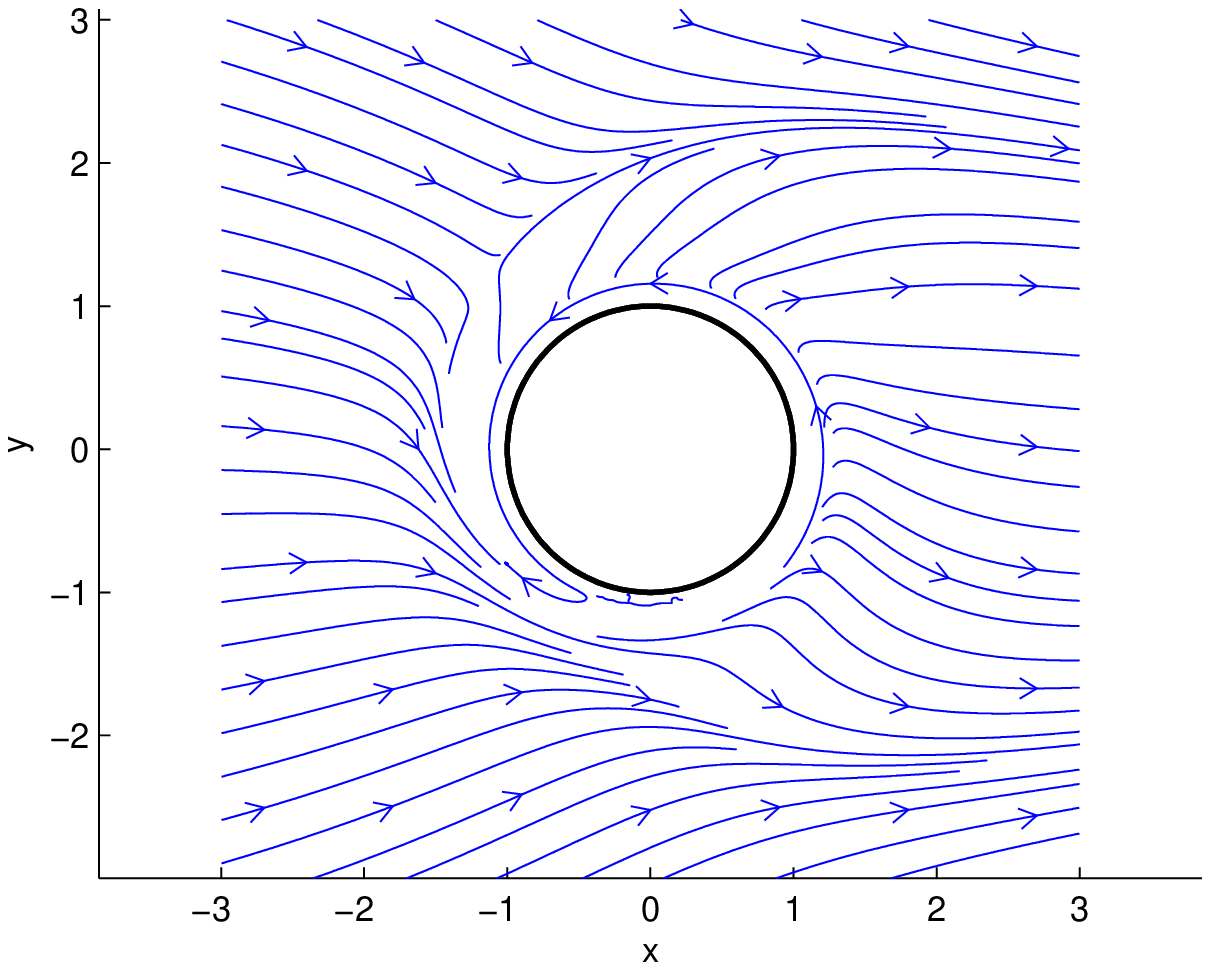}
\caption{The magnetic lines for a maximally rotating BH in 
uniform translational motion in the $y$-direction. Left: the same magnetic
field as in Fig~\ref{fig1} ($a=1$), but with the linear
velocity of the black hole turning the asymptotic direction of
the field lines far from the black hole. The motion also affects the field
structure near the horizon. Right: 
the same case as in the left panel, but for the counter-rotating BH ($a=-1$).
\label{fig2}}
\end{figure*}

\begin{figure*}[tbh!]
\includegraphics[width=.49\textwidth]{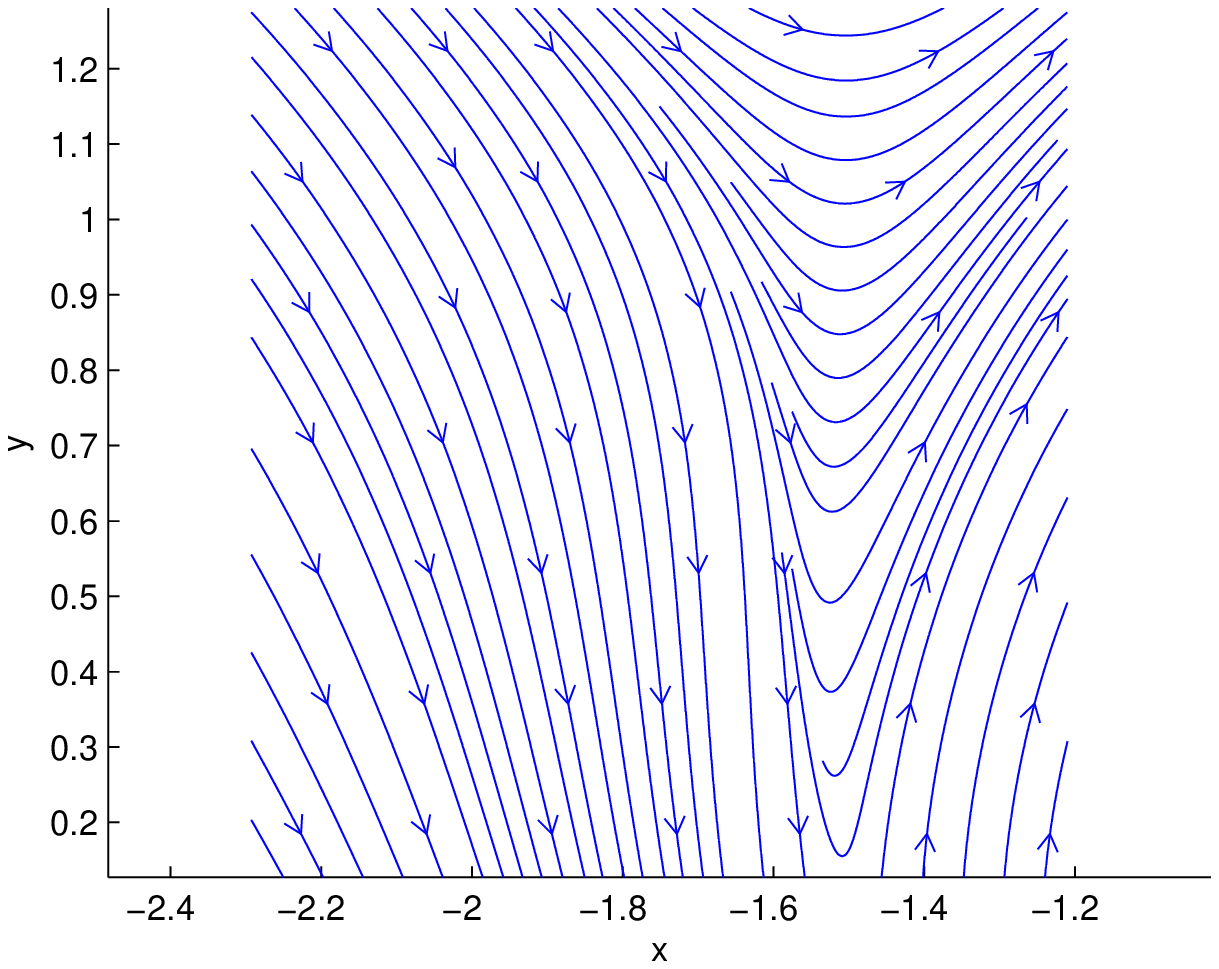}
\includegraphics[width=.49\textwidth]{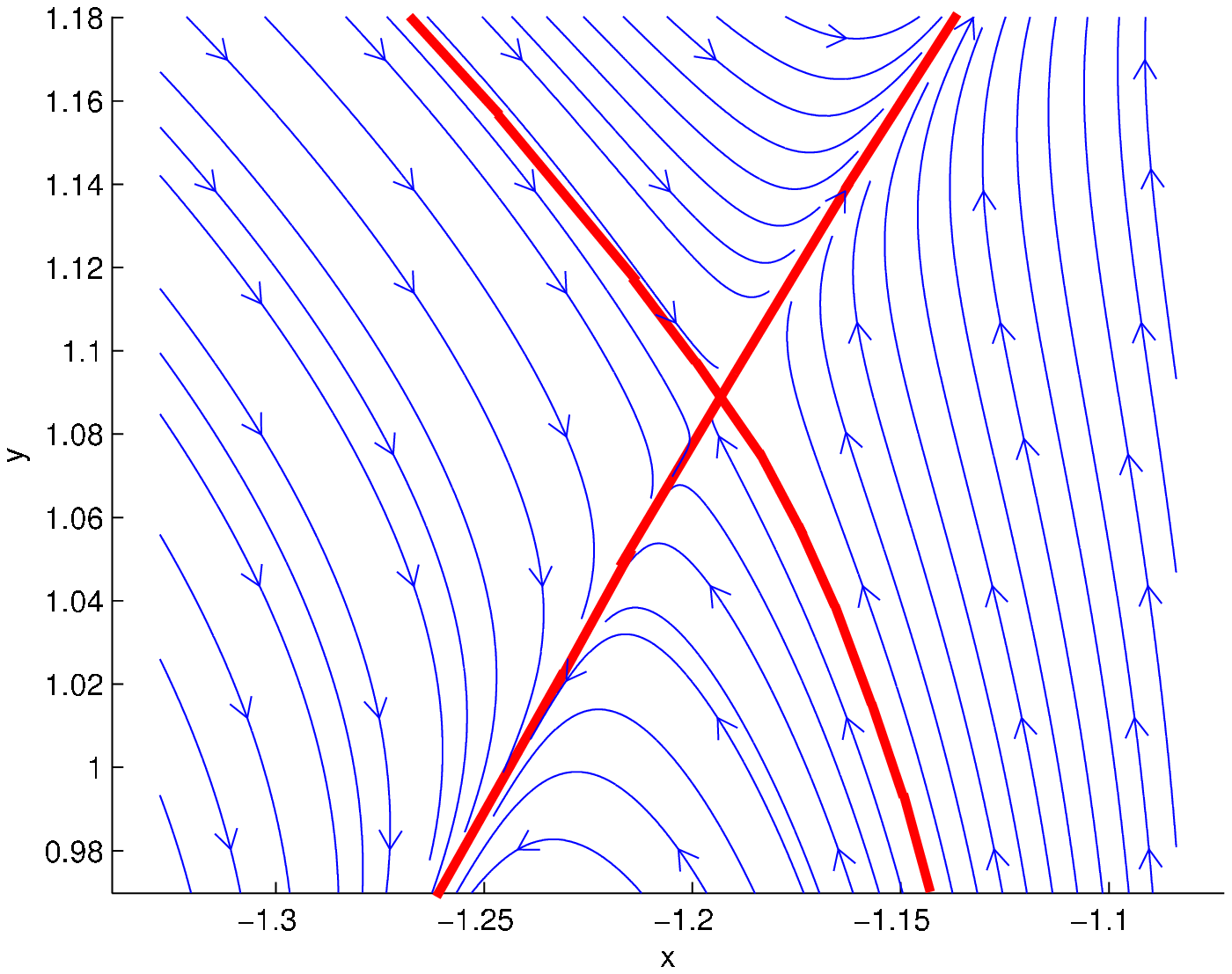}
\caption{The layered structure of the field lines near the horizon.
Left: a detail of the magnetic field from the left panel of
Fig~\ref{fig1}. Right: a detail of the right panel of Fig~\ref{fig2}. The
separatrix curve (red) is clearly revealed around the magnetic null
point.
\label{fig3}}
\end{figure*}

\section{Magnetic layers and null points}
Figure~\ref{fig1} shows the typical structure of the field lines
representing an asymptotically transverse field ($B_{\parallel}=0$).
Albeit we have fixed the boundary condition corresponding to the
homogeneous field outside the ergosphere, its near-horizon structure is
much more dramatic. In fact, it is sensitive to the direction of the
field with respect to the rotation axis. Formation of the layers in the
ergospheric region is an interesting feature of the rotating spacetime.
Two critical points can be seen occurring at radii up to $r\simeq1.7$ for
$a=1$. Fig.~\ref{fig1} shows two plots that differ from each other only
by the sense of the BH rotation.

As mentioned above, we construct the field lines with respect to the
frame orbiting freely in the equatorial plane. The two panels shown in
Fig.~\ref{fig1} are not merely a symmetrical inversions of each other,
and this asymmetry is indeed due to the frame motion, which is oriented
in the same (positive) sense in both examples. Nonetheless, the
alternating direction of the field lines arises in both cases. We notice
that the gravito-magnetically induced electric field does not vanish
along the magnetic lines, and it is thus capable of accelerating the
charged particles.

In the outlined model we have imagined the origin of the magnetic field
by amplification in the accreted plasma. However, recent studies in
numerical relativity point to another possibility, namely, that a
supermassive BH receives a high kick velocity. Recently,
\citet{clz07} and \citet{ghs07} came to the conclusion that a black hole
could gain a recoil velocity up to $\simeq0.5$ per cent of the speed of
light as a result of an anisotropic emission of gravitational waves
during the merger process. The merged BH is then ejected from the
nucleus of a host galaxy \citep{kl08,km08}. On its way out, the BH
encounters regions where the interstellar magnetic field is enhanced
\citep[the evidence for the localised magnetic field enhancements has
been discussed e.g.\ in the case of Galaxy center BH
magnetosphere;][]{m06}. In order to describe the interacting magnetic
field in such circumstances, one needs to take the BH linear velocity
into account.

So far we neglected any translational motion of the BH. However, a
fast-moving BHs can be included by generalising eqs.\ 
(\ref{mf1})--(\ref{mf4}) in a straightforward manner, allowing us to
construct the magnetic field lines of the misaligned field near a BH in
uniform motion. This motion can be taken into account by the Lorentz
boost of the field intensities $E_{(a)}$, $B_{(a)}$. To this end we
notice that the Lorentz transform, when applied in the asymptotically
distant region, has two consequences on the field components: it
(i)~turns the direction of the uniform magnetic field, and
(ii)~generates a new electric component. 

The electromagnetic test field near a moving black hole,
$\bm{F^\prime}_{\!\!\!r\rightarrow\infty}=\bm{\Lambda_\beta}^{\!\!\!\!\!\rm
T}\bm{F}\bm{\Lambda_\beta}$, is obtained as a superposition of two parts
combined together in the due ratio, i.e.\ the asymptotically uniform
magnetic field, rotated into the desired direction, and the solution for
the asymptotically uniform electric field ($\bm{\Lambda_\beta}$ denotes
the matrix of Lorentz boost to the desired velocity $\beta\equiv v/c$).
The form of the latter is found by applying the dual transform to Wald's
field, which interchanges the role of magnetic and electric terms.
Albeit these are lengthy expressions, the field components with respect
to the moving frame and the tetrad components $\bm{e}_{(a)}^{\mu}$ can
be written in the explicit form.\footnote{Just note that
this approach ignores any back-reaction that the electromagnetic field
may exercise on the black hole motion and rotation. However, this
higher-order effect acts on very long time-scales \citep{gap84}, so we
can neglect it.} 
We include the BH linear motion in Figure~\ref{fig2}. In order to
show the effect clearly, we choose high velocity, $\beta_y=-0.99$. 
Let us note that
the tightly layered structure of the magnetic field lines becomes quite
confusing very near the horizon. Alternatively, it can be plotted 
in terms of a new radial
variable, $r^*\equiv1-1/r$.
It allows us to
stretch the complicated structure caused by the
frame-dragging just above the horizon, as the latter is brought to the
coordinate system origin \citep[for these plots and for the
explicit form of the field components, see][]{k08}. 

Figure~\ref{fig3} shows a detail of the magnetic structure. As one
proceeds towards the horizon, the magnetic layers become progressively 
narrow and eventually develop a magnetic null point. The most
interesting case is shown in the right panel, where we capture a
separatrix curve, distinguishing the location of the null point well
above the horizon. For the latter plot we again set a constant velocity
directed along the $y$-axis. The null point arises just above ISCO,
located at $x^2+y^2=1$ for $a=1$. Translational velocity helps to
stretch the magnetic lines further out from the horizon. We find that
the null point can be as far as one gravitational radius from the
horizon. The offset increases with $a$ and is also influenced by the
black hole motion.

The electric field is induced by the magnetic component, and in this
sense the solution is self-consistent. Particles are accelerated very
efficiently by the electric field that remains non-vanishing across the
neutral point. We do not show the associated electric field just because
its lines do not reside in a single plane, and so the structure of the
electric field is more difficult to visualise \citep[see][]{k08}.

We expect reconnection to occur intermittently, as more plasma is
injected into the dissipation region or created via the pair-cascades and 
two-photon reactions. The presumed source of additional material is the
accretion disc truncated above the ISCO, or passing stars that are
damaged by tidal forces. On the other hand, the physical origin of
dissipation is a matter of long-lasting debate. Under the conditions of
a magnetically dominated collisionless medium, the anomalous
reconnection is a promising scenario \citep{v76,bp07}. Reconnection has
been observed also in the simulations of convection dominated flows of
\citet{ina03}. However, in the latter work the MHD effects on the
sheared plasma motion dominate the system evolution, while the
approximation of pseudo-Newtonian gravity does not permit to capture the
frame-dragging effects. 

Very recently, \citet{ka08} suggested that the magnetic reconnection may
indeed be a relevant mechanism for the energy extraction from the
ergosphere. Nevertheless, \citeauthor{ka08} restrict their discussion to
the special-relativistic MHD framework. The formalism of resistive MHD
has been applied to the BH accretion \citep[e.g.,][]{kk96}, and in this
case an interplay between MHD and gravitational effects should have a
significant role. The regime of the intermittent reconnection from the
almost evacuated ergosphere, as envisaged here in our paper, represents
a complementary situation.

\section{Conclusions}
We considered an interesting possibility that the immediate
neighbourhood of a rotating BH may be threaded by an organised component
of the magnetic field, the origin of which is in currents flowing
outside the black hole. Albeit more challenging to recognise than BHs in
a standard accretion regime, the diluted gas in black hole ergospheres
is an interesting possibility in which the turbulent motions are
suppressed by the ordered field. Hence, they could represent a suitable
system to study the relativistic effects in a resistive plasma near the
horizon, such as the Meissner expulsion of the magnetic flux with the
increasing angular momentum. Even if we adopted an idealised system, we
have seen that the magnetic structure becomes very rich just by pure
interaction with the gravito-magnetic field.

We examined the influence of BH rotation acting on the ordered magnetic
field in the physical frame of matter orbiting a black hole, or plunging
down to it. If rotation is fast enough, the magnetic layers and the
corresponding null points exist just above the ISCO. Although we
prescribed a special configuration of the magnetic field, the process
of warping the field lines is a general feature that should operate also
in more complicated settings. The layered structure of the magnetic
field lines with neutral points suggest this should become a site of
particle acceleration.

The essential ingredients of the model are the {\it rotating black hole}
and the {\it oblique magnetic field} in which the BH is embedded. The
interaction region is very near the horizon, representing, to our
knowledge, the acceleration site nearest to the BH horizon among the
variety of mechanisms proposed so far. For a rapidly rotating black hole
the magnetic null point occurs above ISCO. Our results demonstrate that
the gravitational field of a rotating BH creates by itself a very
complicated structure near the horizon. 

The assumed electrovacuum solution (\ref{mf1})--(\ref{mf4}) captures the
essential physical ingredients of our investigation, i.e.\ the organised
field near a rotation black hole. However, it should be quite obvious
that the realistic field structure must differ from the one described in
our paper. Perfectly vacuum black holes are unlikely in any
astrophysical system, as the horizon neighbourhood is expected to be
continuously supplied with electron--positron pairs.

We concentrated on the equatorial plane in which the transverse magnetic
field lines reside, so they can be readily plotted, but this constraint
was imposed only to keep the graphs as clean as possible. In spite of 
simplified geometry, the processes taking place so close to the horizon
are clearly difficult to investigate observationally. Plasma motions
near the BH are currently inaccessible to direct imaging but could be
resolved by interferometric techniques in the future. With a planned
accuracy of 10 micro-arcseconds, GRAVITY will have a capability to carry
out such type of observation in the near infrared domain.

\ack
We thank the Czech Science Foundation (205/07/0052) for a continued
support.


\begin{thebibliography}{}
\bibitem[Albert et al(2008)]{a08}Albert J et al 2008 \textit{ApJL} \textbf{685} L23
\bibitem[Aliev and Galtsov(1989)]{ag89}Aliev A~N, Galtsov D~V 1989 \textit{Sov Phys} \textbf{32} 75
\bibitem[Balbus and Hawley(1998)]{bh98}Balbus S~A and Hawley J~F 1998 \textit{Rev Mod Phys} \textbf{70} 1
\bibitem[Barausse et al(2007)]{bhr07}Barausse E, Hughes S~A, and Rezzolla L 2007 \textit{Phys Rev D} \textbf{76} 044007
\bibitem[Bardeen et al(1972)]{bpt72}Bardeen J~M, Press W~H, and Teukolsky S~A 1972 \textit{ApJ} \textbf{178} 347
\bibitem[Beckwith et al(2008)]{bhk08}Beckwith K, Hawley J~F, and Krolik J~H 2008 \textit{ApJ} \textbf{678} 1180
\bibitem[Begelman et al(1984)]{bbr84}Begelman M~C, Blandford R~D, and Rees M 1984 \textit{Rev Mod Phys} \textbf{56} 255
\bibitem[Bi\v c\'ak and Jani\v s(1980)]{bj80}Bi\v{c}\'ak J and Jani\v{s} V 1980 \textit{MNRAS} \textbf{212} 899
\bibitem[Bi\v{c}\'ak et al(2007)]{bkl07}Bi\v{c}\'ak J, Karas V, and Ledvinka T 2007 in: {\it Black Holes from Stars to Galaxies} Proc of IAU Symp 238 (Cambridge: Cambridge University Press) p~144
\bibitem[Birn and Priest(2007)]{bp07}Birn J and Priest E~R 2007 \textit{Reconnection of Magnetic Fields, Magnetohydrodynamics and Collisionless Theory and Observations} (Cambridge: Cambridge University Press)
\bibitem[Bisnovatyi-Kogan and Lovelace(2007)]{bl07}Bisnovatyi-Kogan G~S and Lovelace R~V~E 2007 \textit{ApJL} \textbf{667} L167
\bibitem[Burgarella et al(1993)]{blo93}Burgarella D, Livio M, and O'Dea C~P 1993 \textit{Astrophysical Jets} (Cambridge: Cambridge University Press)
\bibitem[Campanelli et al(2007)]{clz07}Campanelli M, Lousto C, Zlochower Y, and Merritt D 2007 \textit{ApJL} \textbf{659} L5
\bibitem[Coker and Melia(1997)]{cm97}Coker R~F and Melia F 1997, \textit{ApJL} \textbf{488} L149
\bibitem[Cuadra et al(2008)]{cnm08}Cuadra J, Nayakshin S, and Martins F 2008 \textit{MNRAS} \textbf{383} 458
\bibitem[Doeleman et al(2008)]{dwr08}Doeleman S~S, Weintroub J, Rogers A~E~E et al 2008 \textit{Nature} in press
\bibitem[Eckart et al(2008)]{ebz08}Eckart A, Baganoff F~K, Zamaninasab M et al 2008 \textit{A\&A} \textbf{479} 625
\bibitem[Fendt and Greiner(2001)]{fg01}Fendt C and Greiner J 2001 \textit{A\&A} \textbf{369} 308
\bibitem[Galtsov et al(1984)]{gap84}Galtsov D~V, Aliev A~N, and Petukhov V~I 1984 \textit{Phys Lett} \textbf{105A} 346
\bibitem[Gonz\'alez et al(2007)]{ghs07}Gonz{\'a}lez J~A, Hannam M, Sperhake U, Br{\"u}gmann B, and Husa S 2007 \textit{Phys Rev Lett} \textbf{98} 231101
\bibitem[Haugen and Brandenburg(2004)]{hb04}Haugen N~E~L and Brandenburg A 2004 \textit{Phys Rev E} \textbf{70} 036408
\bibitem[Igumeshchev et al(2003)]{ina03}Igumenshchev I~V, Narayan R, and Abramowicz M~A 2003 \textit{ApJ} \textbf{592} 1042
\bibitem[Junor et al(1999)]{jbl99}Junor W, Biretta J~A, Livio M 1999 \textit{Nature} \textbf{401} 891
\bibitem[Karas(1989)]{k89}Karas V 1989 \textit{Phys Rev D} \textbf{40}, 2121
\bibitem[Karas and Vokrouhlick\'y(1994)]{kv94}Karas V and Vokrouhlick\'y D 1994 \textit{ApJ} 422 208
\bibitem[King et al(1975)]{kkl75}King A~R, Kundt W, and Lasota J~P 1975 \textit{Phys Rev D} \textbf{12} 3037
\bibitem[Koide(2004)]{k04}Koide S 2004 \textit{ApJL} \textbf{606} L45
\bibitem[Koide and Arai(2008)]{ka08}Koide S and Arai K 2008 \textit{ApJ} \textbf{682} 1124
\bibitem[Koide et al(2006)]{kks06}Koide S, Kudoh T, and Shibata K 2006 \textit{Phys Rev D} \textbf{74} 044005
\bibitem[Komissarov et al(2007)]{kbv07}Komissarov S~S, Barkov M~V, Vlahakis N, and K\"onigl A 2007 \textit{MNRAS} 380 51
\bibitem[Komissarov and McKinney(2007)]{km07}Komissarov S~S and McKinney J~C 2007 \textit{MNRAS} \textbf{377} L49
\bibitem[Komossa and Merritt(2008)]{km08}Komossa S and Merritt D 2008 \textit{ApJL} \textbf{683} L21
\bibitem[Kop\'a\v{c}ek(2008)]{k08}Kop\'a\v{c}ek O 2008 in: \textit{Proc WDS 2008 Conf} (Prague: Charles Univ) in press
\bibitem[Kornreich and Lovelace(2008)]{kl08}Kornreich D~A and Lovelace R~V~E 2008 \textit{ApJ} \textbf{681} 104
\bibitem[Krolik et al(2005)]{khh05}Krolik J~H, Hawley J~F, and Hirose S 2005 \textit{ApJ} \textbf{622} 1008
\bibitem[Kudoh and Kaburaki(1996)]{kk96}Kudoh T and Kaburaki O 1996 \textit{ApJ} \textbf{460} 199
\bibitem[Markoff et al(2001)]{mfy01}Markoff S, Falcke H, Yuan F, and Biermann P~L 2001 \textit{A\&A} \textbf{379} L13
\bibitem[McKinney(2005)]{m05}McKinney J~C 2005 \textit{ApJL} \textbf{630} L5
\bibitem[McKinney and Gammie(2004)]{mg04}McKinney J~C and Gammie C~F 2004 \textit{ApJ} \textbf{611} 977
\bibitem[McKinney and Narayan(2007)]{mn07}McKinney J~C and Narayan R 2007 \textit{MNRAS} \textbf{375} 513
\bibitem[Merloni et al(1999)]{mvs99}Merloni A, Vietri M, Stella L, and Bini D 1999 \textit{MNRAS} \textbf{304} 155
\bibitem[Misner et al(1973)]{mtw73}Misner C~W, Thorne K~S, and Wheeler J~A 1973 \textit{Gravitation} (New York: W~H Freeman \& Co)
\bibitem[Morris(2006)]{m06}Morris M 2006 \textit{J Phys: Conf Series} \textbf{54} 1
\bibitem[Neronov and Aharonian(2007)]{na07}Neronov A and Aharonian F~A 2007 \textit{ApJ} \textbf{671} 85
\bibitem[Priest and Forbes(2000)]{pf00}Priest E and Forbes T 2000 \textit{Magnetic Reconnection} (Cambridge: Cambridge Univ Press)
\bibitem[Proga and Begelman(2003)]{pb03}Proga D and Begelman M~C 2003 \textit{ApJ} \textbf{592} 761
\bibitem[Punsly(2008)]{p08}Punsly B 2008 \textit{Black Hole Gravitohydromagnetics} (New York: Springer)
\bibitem[Rockefeller et al(2005)]{rfm05}Rockefeller G, Fryer C~L, and Melia F 2005 \textit{ApJ} \textbf{635} 336
\bibitem[Rothstein and Lovelace(2008)]{rl08}Rothstein D~M and Lovelace R~V~E 2008 \textit{ApJ} \textbf{677} 1221
\bibitem[Somov(2006)]{som06}Somov B~V 2006 \textit{Plasma Astrophysics}, Astr Sp Sci Lib \textbf{341} (New York: Springer)
\bibitem[Stone and Norman(1992)]{sn92}Stone J~M and Norman M~L 1992 \textit{ApJ} \textbf{389} 297
\bibitem[Tomimatsu(2000)]{t00}Tomimatsu A 2000 \textit{ApJ} 528 972
\bibitem[van Hoven(1976)]{v76}van Hoven G 1976 \textit{Sol Phys} \textbf{49} 95
\bibitem[Wald(1974)]{wal74}Wald R~M 1974 \textit{Phys Rev D} \textbf{10} 1680
\end{thebibliography}
\end{document}